\documentclass[aps,pre,twocolumn,showpacs]{revtex4-1}

\usepackage{amsmath,amssymb,graphicx,graphics,bm}
\usepackage{hyperref}
\usepackage{graphics,graphicx}
\usepackage{amsmath,amssymb}

\begin{document}

\title{Particle-photon radiative interactions and thermalization}

\author{Chiara Pezzotti and Massimiliano Giona}
\email[corresponding author:]{massimiliano.giona@uniroma1.it}
\affiliation{Dipartimento di Ingegneria Chimica, Materiali, Ambiente La Sapienza Universit\`a di Roma\\ Via Eudossiana 18, 00184 Roma, Italy}

\date{\today}

\begin{abstract}
We analyze the statistical properties of radiative transitions for a
molecular system possessing discrete, equally spaced, energy levels, interacting
with thermal radiation at constant temperature. A radiative 
fluctuation-dissipation theorem is derived and the particle 
velocity distribution
analyzed. It is shown analytically that, neglecting molecular
collisions, the velocity distribution function cannot be Gaussian, as the
equilibrium value for the kurtosis $\kappa$ is different from $\kappa=3$.
A Maxwellian velocity distribution can be recovered in the limit of small radiative friction.
\end{abstract}

\maketitle

\section{Introduction}
\label{sec1}
One of the main contribution of quantum mechanics is the discovery that
molecules do possess an internal energy structure, owing to which
they radiatively interact with the electromagnetic field
via emission and absorption of energy quanta (photons) \cite{quantum1,quantum2}.
This phenomenon has a deep influence on the statistical and thermodynamic
properties of molecular systems, as shown by Einstein, Debye and many others
\cite{thermo1,thermo2}, both as regards equilibrium and non-equilibrium properties.
The theory of the specific heats of molecules (for instance diatomic molecules)
\cite{diatomic}
and of solids  \cite{solid1} cannot be correctly framed without considering the quantum
description of the excitations of the internal mechanical degrees of freedom.

 The electromagnetic field, described by the
system of the Maxwell equations, is responsible for mechanical actions 
in its interaction with material bodies (massive matter) due
to momentum exchange (radiation pressure). This is well known 
since Maxwell's times, and this currently finds  important fields of 
applications
in the study of condensed matter, in microtechnology and microfluidics, 
in biology, as it
is possible to manipulate mechanically micrometric particles, cells, and molecules through the use of light beams
 (optical tweezers)  \cite{tweezers1,tweezers2,tweezers3}, focus particles at a given spatial location (optical traps) \cite{traps}, or induce extreme thermal
conditions in molecular assemblies via optical interactions (laser cooling techniques) 	\cite{cooling1,cooling2,cooling3}.

Nevertheless,
 the most remarkable effects, as regards thermodynamic properties, is that
the momentum exchange between matter and radiation (recoil effect) is
the physical mechanism leading to thermalization.
As shown by Einstein  \cite{einstein1916}, considering exclusively radiative
interactions between a molecular gas of identical molecules of mass $m$,
and thermal radiation at constant temperature $T$, the squared variance
of the particle velocity entries $\langle v_i^2 \rangle$, $i=1,2,3$,
(since $\langle v_i \rangle=0$)  equals  at equilibrium the
Maxwellian result \cite{thermo2}
\begin{equation}
\langle v_i^2 \rangle = \frac{k_B \, T}{m}
\label{eq0}
\end{equation}
where $k_B$ is the Boltzmann constant.

In this article we analyze the statistical properties of this interaction,
formulating the problem in the form of a stochastic process over the 
increments of a Poisson counting process, applying the formalism recently proposed in \cite{PG1} for stochastic chemical reactions.
Extending the analysis to momentum  transfer, we derive a radiative fluctuation-dissipation relation and the statistical properties of the  particle
velocities at equilibrium.
Throughout this article we consider
exclusively radiative interactions as regards particle momentum
dynamics, deliberately neglecting the
influence of particle-particle collisions. This choice has been made
in order to enucleate and clarify the effects of the momentum
exchange between particles and radiation on the statistical
mechanical properties of a particle gas. 
Despite the fact
that radiative interactions provide the physical mechanism for thermalization,
 in the
meaning of eq. (\ref{eq0}), the resulting velocity distributions
deviate from the Maxwellian, and a Gaussian shape is recovered in the
limit of small radiative friction.

The article is organized as follows. Section \ref{sec2} briefly introduces
the problems and reviews the basic conservation principles that apply,
and the meaning of Einstein's result \cite{einstein1916}. 
Section \ref{sec3} develops the stochastic equations for the
occupation numbers of the internal energy levels of a molecular system
interacting with a given number of photons. We adopt the approximation
of closed system discussed in \cite{lami1985}, deriving the equilibrium
properties. Section \ref{sec4}  addresses the thermalization problem, i.e.,
the statistics of the momentum exchange between a particle gas and thermal 
radiation at constant temperature $T$. A new stochastic formulation
of the particle equations of motion over the increments of a Poisson process
is   developed (the Appendix addresses some technicalities  associated with
this class of equations). A radiative fluctuation-dissipation theorem
is formulated and the functional form of the velocity
distribution function thoroughly considered in Section \ref{sec6}, 
showing its generic deviation
from the Maxwellian behavior.

\section{Radiative interactions}
\label{sec2}

The interaction between  a molecular system with radiation develops through:
(i) radiative processes of emission and absorption of radiation, (ii)
photon-molecule scattering, (iii) interactions with the 
zero-point energy field \cite{zpf1,zpf2}.
According to the analysis developed in \cite{einstein1916}, we neglect scattering processes, and the interactions  with zero-point fluctuations, focusing exclusively
on  the effects of radiative transitions. 
Consider the transition of a quantum system (molecule) 
from the energy level $E_1$ to the energy level $E_2$ due to
absorption of an energy quantum of frequency $\nu$, with
\begin{equation}
E_2-E_1 =h \, \nu
\label{eq_trr1}
\end{equation}
where $h$ is the Planck constant. This elementary event fulfils the fundamental principles of conservation
of energy and momentum. Let  ${\bf v}_1$ and ${\bf v}_2$ be the velocities of the molecule before and after the
radiative interaction with the photon (in the present case  an absorption event). Since a photon of energy $h \, \nu$ possesses
a momentum ${\bf p}_\phi$ given by
\begin{equation}
{\bf p}_\phi= \frac{ h \, \nu}{c} \, {\bf n}
\label{eq_trr2}
\end{equation}
where $c$ is the speed of light {\em in vacuo} and ${\bf n}$ the unit vector in the direction of
propagation, in the low-velocity limit, (so that relativistic corrections can be neglected),  the energy balance reads
\begin{equation}
E_1 + \frac{1}{2} m \, |{\bf v}_1|^2 + h \, \nu = E_2 + \frac{1}{2} \, m \, |{\bf v}_2|^2
\label{eq_trr3}
\end{equation}
where $m$ is the mass of the molecule, and the momentum balance takes the form
\begin{equation}
m \, {\bf v}_1 + \frac{h \, \nu}{c} \, {\bf n} = m \, {\bf v}_2
\label{eq_trr4}
\end{equation}
As the kinetic energy contributions are negligible, 
since using eq. (\ref{eq_trr4}), eq. (\ref{eq_trr3}) can be expressed
as
\begin{equation}
E_1+ h \, \nu \, (1-\varepsilon)= E_2 \, , \qquad
\varepsilon = \frac{2 \, {\bf v} \cdot {\bf n}}{c}+ \frac{h\, \nu}{m \, c^2}
\label{eq_trr4bis}
\end{equation}
and $\varepsilon$ is small in the non-relativistic limit ($v/c \ll 1$), and for
generic molecular systems ($h \nu/m c^2 \ll 1$),
eq. (\ref{eq_trr3}) can be simplified as
\begin{equation}
E_1 + h \, \nu = E_2
\label{eq_trr5}
\end{equation}
As observed in \cite{einstein1916}, the radiative interactions, once 
 considered in the
reference frame of the moving particle, should account for relativistic 
corrections,  specifically  related to the property that
 the equilibrium spectral density of the radiation (the Planck distribution) is not Lorentz covariant. Once
expressed  in the reference frame of the molecule involved in the interaction,
 a dissipartive term , proportional to the ratio ${\bf v}_1/c$
arises in the momentum balance, so that eq. (\ref{eq_trr4}) should be substituted by a dissipative
dynamics, containing a friction term, the occurrence of which  provides
the equilibrium result eq. (\ref{eq0}), see also \cite{thermo2,milonni}.
In the remainder we take this result for given, leaving to a future
work a  thorough  and comprehensive discussion on the validity of momentum
conservation in radiative processes.

\section{Thermalization of internal quantum states}
\label{sec3}

Consider a system of identical molecules interacting with radiation through emission and absorption
of energy quanta. For simplifying the analysis, the following assumptions are made:
\begin{itemize}
\item the molecules are characterized by a countable number of  equally spaced  energy  levels $E_k$, with
$E_{k+1} > E_k$, $k=0,1,\dots$, and 
\begin{equation}
E_{k+1}-E_k = E_\delta = h \, \nu
\label{eq_trr6}
\end{equation}
\item the molecules interact with a photon gas at thermal equilibrium with temperature $T$;
\item the system is closed and isolated, both as regards molecules and photons.
\end{itemize}
The latter assumption simplifies the analysis but does not alter the
physics of the problem and the results obtained at equilibrium.

Let $p_k(t)$ be the number density in the occupation of the $k$-level at time $t$,  $p_k^0$ its initial value
at time $t=0$,  $q(t)$ the density of photons at the resonant frequency $\nu$, and $q^0$ its initial
value.
Mass conservation implies that
\begin{equation}
\sum_{k=0}^\infty p_k(t)= \sum_{k=0}^\infty p_k^0 = P_{\rm tot}
\label{eq_trr7}
\end{equation}
As the system is supposed to be isolated, i.e. closed with respect to  radiation, the principle of conservation
of the ``virtual photon number'' applies, dictating that,
\begin{equation}
q(t)+\sum_{k=1}^\infty  k \, p_h(t)= q^0+\sum_{k=0}^\infty  k \, p_k^0
\label{eq_trr8}
\end{equation}
The interaction of the molecules with the photon gas  implies the emission/absorption of radiation, where the emission process
can be either a spontaneous or  a stimulated transition, i.e. 
induced by a collision with an incoming photon.
Therefore, if $\lambda$ is the emission rate, we have
\begin{equation}
\lambda= \lambda_s + \lambda_0 \, q(t)
\label{eq_trr9}
\end{equation}
while the absorption rate $\mu$ is given by
\begin{equation}
\mu = \mu_0 \, q(t)
\label{eq_trr10}
\end{equation}
As shown by Einstein \cite{einstein1916}, the specific rate of absorption and stimulated emission should be equal, i.e.,
\begin{equation}
\mu_0=\lambda_0
\label{eq_trr11}
\end{equation}
Consider for simplicity transition processes betwen nearest nighbouring energy level. The inclusion
of higher-order transitions does not add any new physics, making solely the notation  more complicated and lengthy.
The balance equations for this process read
\begin{eqnarray}
\frac{d p_k(t)}{d t}  & =  & - \left [(\lambda_s + \lambda_0 q(t) ) \, \eta_k + \lambda_0 \, q(t) \right ] \, p_k(t) \nonumber \\
&+&
(\lambda_s+\lambda_0 \, q(t)) \, p_{k+1}(t)+ \lambda_0 \, q(t) \, p_{k-1}(t) 
\label{eq_trr12}
\end{eqnarray}
$k=0,1,\dots$, where $\eta_0=0$, $\eta_k=1$ for $k=1,2,\dots$, and $p_{-1}=0$,
and
\begin{equation}
\frac{d q(t)}{d t}= (\lambda_s+\lambda_0 \, q(t)) \sum_{k=1}^\infty p_k(t) - \lambda_0 \, q(t) \sum_{k=0}^\infty p_k(t)
\label{eq_trr13}
\end{equation}
In a stochastic representation of the process, let $N_k(t)$ be the number of molecules in the $k$-th level,
and $N^q(t)$ the photon number. If $N_g$ is the granularity number chosen \cite{PG1}, $\sum_{k=0}^\infty N_k(0)=N_g$,
the relations between $N_k(t)$ and $p_k(t)$ and between $q(t)$ and $N^q(t)$ are expressed 
by
\begin{equation}
p_k(t)= P_{\rm tot} \, \frac{N_k(t)}{N_g} \, , \qquad q(t) = P_{\rm tot} \, \frac{N^q(t)}{N_g}
\label{eq_trr14}
\end{equation}
Expressed in terms of $N_k(t)$, $N^q(t)$, the balance equations (\ref{eq_trr12})-(\ref{eq_trr13})
thus become
\begin{eqnarray}
\frac{d N_k(t)}{d t} & =  &- \left [(\widetilde{\lambda}_s + \widetilde{\lambda}_0 N^q(t) ) \, \eta_k + \widetilde{\lambda}_0 \, N^q(t) \right ] \, N_k(t) \nonumber \\
&+ &
(\widetilde{\lambda}_s+\widetilde{\lambda}_0 \, N^q(t)) \, N_{k+1}(t)+ \widetilde{\lambda}_0 \, N^q(t) \, N_{k-1}(t) \nonumber \\
\frac{d N^q(t)}{d t} & =  & (\widetilde{\lambda}_s+\widetilde{\lambda}_0 \, N^q(t)) \sum_{k=1}^\infty N_k(t) - \widetilde{\lambda}_0 \, N^q(t) \sum_{k=0}^\infty N_k(t) \nonumber \\
\label{eq_trr15}
\end{eqnarray}
with
\begin{equation}
\widetilde{\lambda}_s=\lambda_s \, , \qquad \widetilde{\lambda}_0= \lambda_0 \, \frac{P_{\rm tot}}{N_g}
\label{eq_trr16}
\end{equation}
A stochastic  Markovian dynamics follows  from eqs. (\ref{eq_trr15})-(\ref{eq_trr16}), applying the formalism developed in \cite{PG1}, by considering as  
stochastic variables the energy state of each molecule and $N^q(t)$.
Let $\sigma_\alpha(t)=0,1,\dots$ be the energy state of the  $\alpha$-th
molecule at time $t$, $h=1,\dots,N_g$. The evolution of $\{\sigma_\alpha(t) \}_{\alpha=1}^{N_g}$ follows the Markovian dynamics,
\begin{equation}
\frac{d \sigma_\alpha(t)}{d t}= - \eta_{\sigma_\alpha(t)}  \frac{d \chi_\alpha^{(e)}(t,\widetilde{\lambda}_s+ \widetilde{\lambda}_0 \, N^q(t))}{d t}+
\frac{d \chi_\alpha^{(a)}(t,\widetilde{\lambda}_0 \, N^q(t))}{d t}
\label{eq_trr17}
\end{equation}
while
\begin{eqnarray}
\frac{d N^q(t)}{d t}  &= & \sum_{\alpha= 1}^{N_g} \, \eta_{\sigma_\alpha(t)} \, \frac{d \chi_\alpha^{(e)}(t,\widetilde{\lambda}_s+ \widetilde{\lambda}_0 \, N^q(t))}{d t} \nonumber \\
&-& \sum_{\alpha=1}^{N_g} \frac{d \chi_\alpha^{(a)}(t,\widetilde{\lambda}_0 \, N^q(t))}{d t}
\label{eq_trr18}
\end{eqnarray}
$\{ \chi_\alpha^{(e)}(t,\widetilde{\lambda}_s+ \widetilde{\lambda}_0 \, N^q(t)) \}_{\alpha=1}^{N_g}$
and $\{ \chi_\alpha^{(a)}(t,\widetilde{\lambda}_0 \, N^q(t)) \}_{\alpha=1}^{N_g}$ being two families of  $N_g$ independent Poisson
counting processes, mutually independent of each other, associated with the emission and absorption events of the
$\alpha$-th molecules. Observe  that the transition rates of these 
processes depend explictly on the photon number $N^q(t)$.

Given $\{\sigma_\alpha(t) \}_{\alpha=1}^{N_g}$, the occupation number $N_k(t)$ of the $k$-th energy
level is given by
\begin{equation}
N_k(t)= \sum_{\alpha=1}^{N_g} \delta_{k,\sigma_\alpha(t)}
\label{eq_trr19}
\end{equation}
where $\delta_{k, \sigma_{\alpha}}$ are the Kronecker symbols, so that $\delta_{k,\sigma_\alpha(t)}=1$ if $\sigma_\alpha(t)=h$, and
zero otherwise.

\begin{figure}
\begin{center}
\includegraphics[width=8cm]{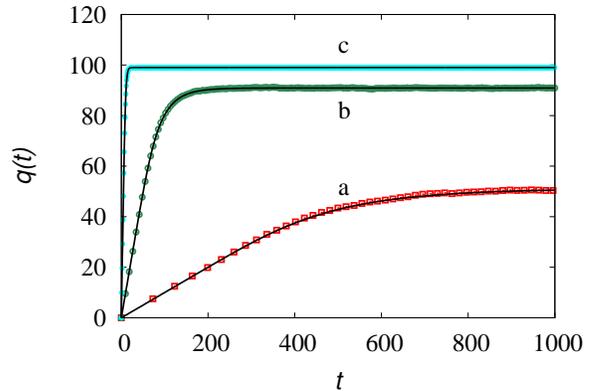}
\end{center}
\caption{$q(t)$ vs $t$ for different values of $\lambda_s$, symbols are the
results of stochastic simulations eq. (\ref{eq_trr17})-(\ref{eq_trr18}), lines
correspond to the solution of the continuous model. Line (a): $\lambda_s=0.01$,
line (b): $\lambda_s=0.1$, line (c): $\lambda_s=1$.}
\label{Fig_pha1}
\end{figure}

\begin{figure}
\begin{center}
\includegraphics[width=8cm]{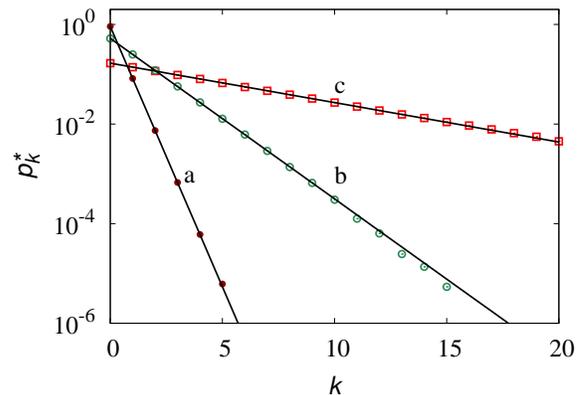}
\end{center}
\caption{$p_k^*$ vs $k$ for different values of $\lambda_s$, symbols are the
results of stochastic simulations eq. (\ref{eq_trr17})-(\ref{eq_trr18}), lines
correspond to the exponential (Boltzmann) distribution (\ref{eq_trr23})
with $q^*$ given by eq. (\ref{eq_trr35}). Line (a): $\lambda_s=0.01$,
line (b): $\lambda_s=0.1$, line (c): $\lambda_s=1$.}
\label{Fig_pha2}
\end{figure}

\begin{figure}
\begin{center}
\includegraphics[width=8cm]{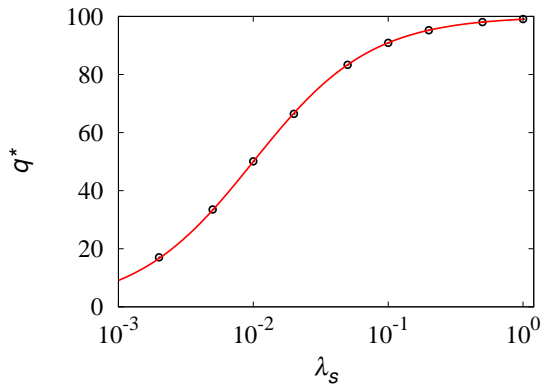}
\end{center}
\caption{$q^*$ vs $\lambda_s$ for $\lambda_0=10^{-3}$, $P_{\rm tot}=10$, $\phi_0=100$.
Symbols ($\circ$) are the results of the stochastic simulations (with $N_g=10^4$),
line corresponds to eq. (\ref{eq_trr35}).}
\label{Fig_pha3}
\end{figure}

\begin{figure}
\begin{center}
\includegraphics[width=8cm]{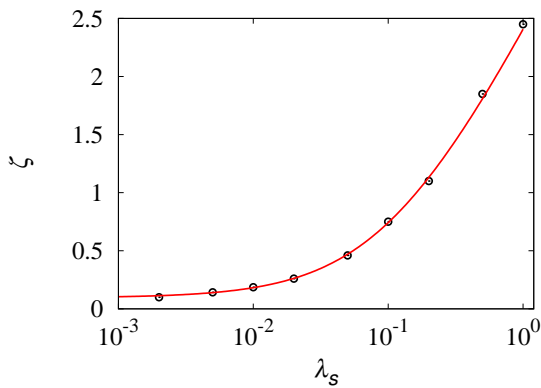}
\end{center}
\caption{Exponent $\zeta$ vs $\lambda_s$ for $\lambda_0=10^{-3}$, $P_{\rm tot}=10$, $\phi_0=100$.
Symbols ($\circ$) are the results of the stochastic simulations (with $N_g=10^4$),
line corresponds to  the theoretical expression obtained from the continuous model.}
\label{Fig_pha4}
\end{figure}

Consider the equilibrium properties of this system, indicating with $p_k^*$ and $q^*$ the equilibrium
values.  Eq. (\ref{eq_trr12}) for $k=0$ at steady state becomes
\begin{equation}
-\lambda_0 \, q^* \, p_0^* + (\lambda_s + \lambda_0 \, q^* ) \, p_1^* =0
\label{eq_trr20}
\end{equation}
so that
\begin{equation}
p_1^* = \frac{\lambda_0 \, q^*}{\lambda_s + \lambda_0 \, q^*} \, p_0^*
\label{eq_trr21}
\end{equation}
An analogous relation applies for generic $k=1,2,\dots$, namely
\begin{equation}
p_{k+1}^* = \frac{\lambda_0 \, q^*}{\lambda_s + \lambda_0 \, q^*} \, p_k^*
\label{eq_trr22}
\end{equation}
Therefore, as expected, the equilibrium distribution of level occupation is
given by
\begin{equation}
p_h = C \, \left ( \frac{\lambda_0 \, q^*}{\lambda_s + \lambda_0 \, q^*} \right )^h = C \,  \exp \left [- h \, \log
\left ( \frac{\lambda_s + \lambda_0 \, q^*}{\lambda_0 \, q^*} \right ) \right ]
\label{eq_trr23}
\end{equation}
It is a discrete Boltzmann distribution, where the term $\log((\lambda_s + \lambda_0 \, q^*)/\lambda_0 \, q^*)$ can
be identified with the Boltzmann factor $E_\delta/k_B \, T$,
\begin{equation}
\frac{E_\delta}{k_B \, T}= \log
\left ( \frac{\lambda_s + \lambda_0 \, q^*}{\lambda_0 \, q^*} \right )
\label{eq_trr24}
\end{equation}
and this provides an alternative definition of equilibrium temperature $T$ based on radiative
interactions
\begin{equation}
T= \frac{E_\delta}{k_B} \frac{1}{\log
\left ( \frac{\lambda_s + \lambda_0 \, q^*}{\lambda_0 \, q^*} \right )}
\label{eq_trr25}
\end{equation}
From eq. (\ref{eq_trr25}),  temperature is uniquely specified, once 
the equilibrium photon density $q^*$ is given. Consequently
for radiative processes the  equilibrium temperature is one-to-one with the 
steady-state value of the photon density $q^*$.
Two limit cases can be considered.
For $\lambda_s \gg \lambda_0 \, q^*$, i.e. for low photon densities, eq. (\ref{eq_trr24}) simplifies
as
\begin{equation}
\log \left ( \frac{\lambda_s}{\lambda_0 \, q^*} \right ) = \frac{E_\delta}{k_B \, T} \,,  \quad \Rightarrow \quad
q^* = \frac{\lambda_s}{\lambda_0} \, e^{-h \, \nu/k_B \, T}
\label{eq_trr26}
\end{equation}
In the opposite case, $\lambda_s \ll \lambda_0 \, q^*$, i.e., in the high photon-density limit,
\begin{equation}
\log \left ( 1+ \frac{\lambda_s}{\lambda_0 \, q^*} \right ) \simeq \frac{\lambda_s}{\lambda_0 \, q^*}= \frac{E_\delta}{k_B \, T}
\label{eq_trr27}
\end{equation}
and thus the equilibrium photon density $q^*$ is proportional to
the temperature $T$
\begin{equation}
q^* = \frac{k_B \, T}{h \, \nu} \, \frac{\lambda_s}{\lambda_0}
\label{eq_trr28}
\end{equation}
Next, consider the expression for $q^*$. Set $\lambda=\lambda_s+\lambda_0 \,q^*$, and $\mu=\lambda_0 \, q^*$, for
notational simplicity, and $x= \mu/\lambda < 1$, so that the equilibrium occupational distribution
can be expressed compactly as $p_k^*= C \, x^k$.
The conditions  expressed by eqs. (\ref{eq_trr7})-(\ref{eq_trr8})  become
at equilibrium
\begin{equation}
\sum_{k=0}^\infty p_k^* = P_{\rm tot}
\label{eq_trr29}
\end{equation}
and
\begin{equation}
q^* + \sum_{k=1}^\infty h \, p_k^* = q^0+ \sum_{k=1}^\infty p_k^0= \Phi_0
\label{eq_trr30}
\end{equation}
Since
$\sum_{k=0}^\infty x^k = \frac{1}{1-x}$, 
$\sum_{k=1}^\infty k \, x^k = \frac{x}{(1-x)^2}$,
we have for the normalization constant $C$,
\begin{equation}
C= P_{\rm tot} \, (1-x)
\label{eq_trr32}
\end{equation}
and  eq. (\ref{eq_trr30}) becomes
\begin{equation}
\Phi_0 - q^* = P_{\rm tot} \, \frac{x}{1-x}
\label{eq_trr34}
\end{equation}
that  can be explicited with respect to $q^*$ to provide
\begin{equation}
q^*= \frac{\Phi_0}{1+ \frac{\lambda_0}{\lambda_s} \, P_{\rm tot}}
\label{eq_trr35}
\end{equation}
Figure 	\ref{Fig_pha1} depicts the evolution of $q(t)$ obtained from
stochastic simulations at $\lambda_0=10^{-3}$ for different values of $\lambda_s$.
In the simulation we have chosen $P_{\rm tot}=10$, and the initial conditions are $p_k^0= 10 \, \delta_{k,10}$,
corresponding to an initial population in the 10-th excited state. 
The stochastic simulation
refers to a granularity number $N_g=10^4$, and $100$ energy levels have been considered.
The steady-state  distributions of the occupation of the energy levels are depicted
in figure \ref{Fig_pha2} for the same values of the parameters of figure \ref{Fig_pha1}.
From the long-term behavior of $q(t)$, the equilibrium value $q^*$ can be obtained. This
is depicted in figure \ref{Fig_pha3} as a function of $\lambda_s$.
Finally, figure \ref{Fig_pha4} depicts the value of the scaling exponent $\zeta$ of $p_k^*$,
$p_k^*= C e^{-k \zeta}$ obtained for the data of figure \ref{Fig_pha2}, compared to the theoretical
expression $\zeta=\log((\lambda_s+\lambda_0 \, q^*)/\lambda_0 \, q^*)$, revealing the
excellent agreement of the stochastic simulations with the theoretical values.

\section{Implications of radiative events in the velocity statistics}
\label{sec4}

From the work by Einstein on emission and absorption of radiation \cite{einstein1916} it becomes
clear that thermalization processes, i.e. the relaxation of  a physical system far from
 equilibrium towards the thermal equilibrium, could be considered as quantum effects
driven by emission and absorption of energy quanta.

This is certainly the case of a diluted gas of massive particles (molecules) interacting with thermal
radiation (i.e. a photon gas at thermal equilibrium, the statistical properties
of which are described by the Planck distribution), in which the following assumptions
can be made:
\begin{itemize}
\item particle dynamics is characterized by two main interactions: (i) emission and absorption
of energy quanta by a particle, and (ii) particle-particle collisions. The assumption of ``diluted
system'' indicates that solely binary collisions are relevant;
\item these two processes can be considered as instantaneous events characterized by a Markovian
transition structure;
\item between two subsequent events (be them particle/photon radiative interactions or particle/particle
collisions), the particle motion is purely inertial, i.e. frictionless and in the absence of external
or interparticle potentials.
\end{itemize}

Moreover, relativistic corrections determines the emergence of a dissipative term in the momentum dynamics proportional to the velocity of the molecule.
Let us further assume that the particles (molecules) can be represented
by a two-level system, where $E_1$ and $E_2$ are the two energy levels $E_2 > E_1$
and $E_2-E_1=h \, \nu$.

In this Section we consider exclusively particle/photon radiative interactions and their effects on
momentum transfer and velocity statistics, leaving the interplay between  radiative processes and
mechanical collisions to a subsequent analysis.

Following the assumptions discussed above, the momentum equation for a generic
particle, due to  the radiative  interactions, can be described by means of the stochastic
differential equation
\begin{equation}
m \, \frac{d {\bf v}}{d t}= \left (-\eta \, {\bf v} + {\bf b} \right ) \, \frac{d \chi(t,\lambda)}{ d t}
\label{eq_ph1}
\end{equation}
where $m$ is the particle mass, and ${\bf v}$ its velocity vector.
 The coefficient $\eta$ is
the {\em radiative friction factor}, possessing the dimension of a mass.
As addressed in Einstein's
work \cite{einstein1916}, the radiative friction is an emergent
property of the momentum exchange between a molecule and a photon during the
radiative process (be it emission or absorption),  related to the well-known recoil effect.
In eq. (\ref{eq_ph1}), $\chi(t,\lambda)$ is a Poisson process possessing transition rate $\lambda$,
and ${\bf b}$ is a random variable corresponding to the photon momentum.

Eq. (\ref{eq_ph1}) is a nonlinear impulse-driven stochastic 
differential equation \cite{imp1,imp2,imp3,impulsive_critic}. Just because
of the presence of the factor $-\eta \, {\bf v}$ multiplying the distributional derivative of the Poisson counting
process, the proper mathematical setting of this class of equations requires some caution,
as addressed in the Appendix. In point of fact, there is a strong analogy
 between
the  mathematical formalization of this class of impulse-driven stochastic differential
equations and the setting of nonlinear Wiener-driven Langevin equations, 
leading to the Ito, Stratonovich, H\"anggi-Klimontovich
formulations \cite{ito,stoca}.

\section{Momentum transfer and radiative fluctuations-dissipation relations}
\label{sec5}

Eq. (\ref{eq_ph1}) describes the momentum transfer  in a radiative process. 
Let $t=t^*$ be a time instant at which $\chi(t,\lambda)$ exhibits a transition, so that in the
neighbourhood of $t^*$ eq. (\ref{eq_ph1}) is equivalent to
\begin{equation}
m \, \frac{d {\bf v}}{d t}= \left (-\eta \, {\bf v} + {\bf b} \right ) \,  \delta(t-t^*)
\label{eq_ph2}
\end{equation}
Integrating the latter equation between $t^*_-$ and $t^*_+$, and letting
${\bf v}={\bf v}(t_-^*)$, ${\bf v}^\prime={\bf v}^\prime(t_+^*)$, ${\bf b}={\bf b}(t^*)$, we have (see Appendix)
\begin{equation}
{\bf v}^\prime= e^{-\eta/m} \, {\bf v} + \frac{\bf b}{\eta} \, (1- e^{-\eta/m})
\label{eq_ph3}
\end{equation}
$\eta= m \, \gamma$, so that $\gamma$ is nondimensional,   
$\alpha=e^{-\gamma}$,
${\bf b}=b \, {\bf r}$, where $b=h \, \nu/c$, corresponding to 
the norm of the  momentum
of a photon with energy $h \, \nu$, and
${\bf r}$ is a unit random vector, $|{\bf r}|=1$, $\langle {\bf r} \rangle=0$,
where $\langle \cdot \rangle$ is the average with respect to the probability
measure of ${\bf r}$.
Thus,
\begin{equation}
{\bf v}^\prime = \alpha \,  {\bf v} + \frac{b \, {\bf r} \, (1-\alpha)}{\eta }
\label{eq_ph5}
\end{equation}
Since it is reasonable to assume the absence of  correlation  (independence) 
between the particle velocity ${\bf v}$ and the
direction ${\bf r}$ of the incoming/emitted photon (this is certainly true for absorption and spontaneous emission, and
it can be extrapolated also  in the case of stimulated emission since particle velocity and the
direction of the incoming photon are certainly uncorrelated from each other), it follows
from eq. (\ref{eq_ph5}) that
\begin{equation}
\langle |{\bf v}^\prime |^2 \rangle = \alpha^2 \, \langle  | {\bf v} |^2 \rangle + \frac{b^2 \, (1 -\alpha)^2}{ \eta^2}
\label{eq_ph6}
\end{equation}
Enforcing at  thermal equilibrium the condition
$\langle | {\bf v}^\prime |^2 \rangle = \langle | {\bf v} |^2 \rangle = 3 \, k_B \, T/m$, 
 we have
\begin{equation}
3 \, \frac{k_B \, T}{m} \, (1-\alpha^2) = \frac{b^2 \, (1-\alpha)^2}{\eta^2}
\label{eq_ph8}
\end{equation}
Eq. (\ref{eq_ph8}) represents the first radiative fluctuation-dissipation 
relation, connecting the nondimensional
friction factor $\gamma$ to  the equilibrium temperature $T$.

In the limit for $\gamma \ll 1$, $e^{-\gamma} \simeq 1-\gamma$, 
and eq. (\ref{eq_ph8})
reduces to
\begin{equation}
6 \, \frac{k_B \, T}{m} \, \gamma= \frac{b^2}{m^2}  \quad \Rightarrow \quad \gamma= \frac{(h \, \nu)^2}{6 \, m \, c^2 \, k_B \, T}
\label{eq_ph9}
\end{equation}
that, setting $E_\phi=h \, \nu$, $E_0=m \, c^2$, $E_T=k_B \, T$ can be rewritten in a more compact way as
\begin{equation}
\gamma= \frac{E_\phi^2}{6 \, E_0 \, E_T}
\label{eq_ph10}
\end{equation}
Eq. (\ref{eq_ph10}) indicates that, in low-friction limit, the nondimensional radiative friction $\gamma$ is
proportional to the ratio of the squared photon energy to the product of the particle rest energy $E_0$ times
the characteristic thermal energy $E_T$.

The momentum dynamics can be  naturally expressed with respect to the operational time $n=0,1,2,\dots$ corresponding
to the number of radiative events occurred, as
\begin{equation}
{\bf v}_{n+1} = \alpha \, {\bf v}_n + \beta \, {\bf r}_{n+1}
\label{eq_ph11a}
\end{equation}
where $\beta=b \, (1-\alpha)/\eta$, and ${\bf r}_{n+1}$ is a family of vector-valued  independent unit random vectors,
uniformly distributed on the  surface of the unit sphere.
The discrete dynamics eq. (\ref{eq_ph11a}) can be explicited
\begin{equation}
{\bf v}_n = \alpha^n \, {\bf v}_0 +  \beta \, \sum_{j=1}^n \alpha^{n-j} \, {\bf r}_j
\label{eq_ph12}
\end{equation}
In the long-term limit, the first term, namely $\alpha^n \, {\bf v}_0$,
depending
on the initial velocity condition, can be ignored as it decays exponentially to zero, so that
\begin{equation}
{\bf v}_n = \beta \, \sum_{j=1}^n \alpha^{n-j} \, {\bf r}_j
\label{eq_ph13}
\end{equation}
Consider the correlation tensor $\langle {\bf v}_n \otimes {\bf v}_p \rangle$, with $p \leq n$,
\begin{equation}
\langle {\bf v}_n \otimes {\bf v}_p \rangle = \beta^2 \, \sum_{j=1}^n \sum_{k=1}^p \alpha^{n+p-j -k} \langle
{\bf r}_j \otimes {\bf r}_k \rangle 
\label{eq_ph14}
\end{equation}
In  the three-dimensional space, enforcing the independence of 
${\bf r}_j$ and ${\bf r}_k$ for $j \neq k$,
and the uniformity of the distribution on the surface of the unit sphere,
we have
\begin{equation}
\langle
{\bf r}_j \otimes {\bf r}_k \rangle = \frac{\delta_{j,k} \, {\bf I}}{3}
\label{eq_ph15}
\end{equation}
where ${\bf I}$ is the identity matrix, so that eq. (\ref{eq_ph14})  becomes
\begin{equation}
\langle {\bf v}_n \otimes {\bf v}_p \rangle = \frac{\beta^2 \, {\bf I}}{3} \, \alpha^{n+p}  \, 
\sum_{k=1}^p \alpha^{-2 \, k}
\label{eq_ph16}
\end{equation}
Making use of the elementary property
\begin{equation}
\sum_{k=1}^p \alpha^{-2 \, k} = \frac{\alpha^{-2} - \alpha^{-2 \, (p+1)}}{1-\alpha^{-2}}
\label{eq_ph17}
\end{equation}
eq. (\ref{eq_ph16}) can be rewritten as
 \begin{equation}
\langle {\bf v}_n \otimes {\bf v}_p \rangle =  \frac{\widetilde{\beta}^2 \, {\bf I} \, \alpha^{n+p}}{3}
\, \left ( \frac{\alpha^{-2} - \alpha^{-2 \, (p+1)}}{1-\alpha^{-2}} \right )
\label{eq_ph18}
\end{equation}
In the long-term limit $n,p \rightarrow \infty$, we obtain
\begin{equation}
\langle {\bf v}_n \otimes {\bf v}_p \rangle = \frac{\beta^2}{3} \, {\bf I} \,
\frac{\alpha^{n-p}}{1-\alpha^2}= \frac{\beta^2}{3 \, (1- \alpha^2)} \, {\bf I} \,  e^{-(n-p) \, \log(1/\alpha)}
\label{eq_ph19}
\end{equation}
The latter result can be expressed with respect to the physical time $t$, as $(n-p)=t/\langle \tau \rangle$,
where $\langle \tau \rangle=1/\lambda$ corresponds to the mean transition time. This leads to the
expression for the velocity autocorrelation tensor,
\begin{equation}
\langle    {\bf v}(t+\tau) \otimes {\bf v}(\tau) \rangle= \frac{\beta^2 \, e^{-\eta_r \, t/m}}{3 \, (1- \alpha^2)} \, {\bf I} 
\label{eq_ph20}
\end{equation}
corresponding to an exponential decay with time $t$, where the effective
radiative dissipation factor $\eta_r$ is  defined by the relation
\begin{equation}
\eta_r=  m \, \lambda \, \log \left (\frac{1}{\alpha} \right )
\label{eq_ph21}
\end{equation}
The effective diffusivity $D$ can be derived from the extension of the Einstein fluctuation-dissipation
relation to radiative processes, $D \, \eta_r = k_B \, T$, to obtain
\begin{equation}
\frac {k_B \, T}{ m \, D}  = \lambda \, \log \left (\frac{1}{\alpha} \right )
\label{eq_ph22}
\end{equation}
In the limit of small $\gamma \ll 1$, $\alpha=1-\gamma$, and thus
\begin{equation}
D= \frac{k_B \, T}{ m \, \lambda \, \log \left ( \frac{1}{1-\gamma} \right ) }
\label{eq_ph23}
\end{equation}
that can be viewed as the second radiative fluctuation-dissipation relation connecting the effective
diffusivity $D$ to the statistics of radiative events.

\section{Statistical characterization of the velocity distribution function}
\label{phsec4}

In this Section we consider the statistical properties of particle velocities emerging from purely
radiative interactions with an equilibrium photon bath.
To this end, it is convenient to discuss separately the 2d case from the 3d situation.
Rescaling the velocity variables ${\bf v} \mapsto {\bf v}/\sigma_{\rm ph}$, with respect to the variance of the photon 
forcing term $\sigma_{\rm ph}$,
\begin{equation}
\sigma_{\rm ph}= \frac{\beta}{\sqrt{d}}
\label{eq_ph24}
\end{equation}
where $d=2,3$,  the rescaled equation   attains the simple form
\begin{equation}
{\bf v}^\prime= \alpha \, {\bf v} + \sqrt{d} \, {\bf r}
\label{eq_ph25}
\end{equation}
where the random vector ${\bf r}$ is defined,  in 2d,  as
\begin{equation}
{\bf r}= \left (
\begin{array}{c}
\cos \phi \\
\sin \phi 
\end{array}
\right )
\label{eq_ph26}
\end{equation}
with a uniform probability density function $p_\phi(\phi)$ 
\begin{equation}
p_\phi(\phi)=\frac{1}{2 \, \pi} \, , \qquad \phi  \in [0,2 \, \pi]
\label{eq_ph27}
\end{equation}
while in  the 3d case
\begin{equation}
{\bf r}= \left (
\begin{array}{c}
\sin \theta \, \cos \phi \\
\sin \theta \,  \sin \phi  \\
\cos \theta
\end{array}
\right )
\label{eq_ph28}
\end{equation}
with a  joint probability density function $p_{\theta,\phi}(\theta,\phi)$
\begin{equation}
p_{\theta,\phi}(\theta,\phi)= \frac{\sin \theta}{4 \, \pi} \, , \qquad (\theta,\phi) \in [0,\pi] \times [0,2 \pi]
\label{eq_ph29}
\end{equation}
Owing to isotropy, in the 2d case it is sufficient to consider the 1d velocity dynamics
\begin{equation}
v^\prime = \alpha \, v + \sqrt{2} \, \cos \phi \, , \qquad  p_\phi(\phi) = \frac{1}{2 \, \pi} \, , \quad \phi \in [0,2 \, \pi]
\label{eq_ph30}
\end{equation}
and  similarly in the 3d case, the equivalent 1d model becomes
\begin{equation}
v^\prime = \alpha \, v + \sqrt{3} \, \cos \theta \, , \qquad p_\theta(\theta) = \frac{\sin \theta}{2} \, , \quad \theta \in [0,\pi]
\label{eq_ph31}
\end{equation}
To begin with, consider the 2d case, for which
\begin{equation}
\langle \cos^2 \phi \rangle = \frac{1}{2} \, , \qquad \langle \cos^4 \phi \rangle = \frac{3}{8}
\label{eq_ph32}
\end{equation}
Therefore, at equilibrium $\langle v \rangle =0$, and
\begin{equation}
\langle v^2 \rangle = \alpha^2 \, \langle v^2 \rangle +1
\label{eq_ph33}
\end{equation}
i.e.,
\begin{equation}
\langle v^2 \rangle = \frac{1}{1-\alpha^2}
\label{eq_ph34}
\end{equation}
As regards the qualitative statistical properties, the main issue is the deviation
from a Gaussian behavior. For this reason, it is interesting to consider the  fourth-order moment and, out of it,
the kurtosis. Enforcing the independence between
 $v$ and  $\phi$ random variables, the  fourth-order
moment   takes the form
\begin{eqnarray}
\langle v^4 \rangle & = & \langle (\alpha \, v + \sqrt{2} \, \cos \phi)^4 \rangle  
 =  \alpha^4 \, \langle v^4 \rangle  + 6 \, \alpha^2 \, \langle v^2 \rangle + \frac{3}{2}
\label{eq_ph35}
\end{eqnarray}
so that
\begin{equation}
\langle v^4 \rangle = \frac{3 \, (1+ 3 \, \alpha^2)}{2 \, (1- \alpha^2) \, (1-\alpha^4)}
\label{eq_ph36}
\end{equation}
From eqs. (\ref{eq_ph34}), (\ref{eq_ph36}) the  expression for the 
kurtosis $\kappa(\alpha)$ follows,
\begin{equation}
\kappa(\alpha)= \frac{\langle v^4 \rangle}{\langle v^2 \rangle^2} =
\frac{3 \, (1+ 3 \, \alpha^2) \, (1-\alpha^2)^2}{2 \, (1- \alpha^2) \, (1-\alpha^4)} 
= \frac{3 \, (1+ 3 \, \alpha^2)}{2 \, (1+\alpha^2)}
\label{eq_ph37}
\end{equation}
A Gaussian behavior is expected for $\alpha \rightarrow 1$, since
\begin{equation}
\lim_{\alpha \rightarrow 1} \kappa(\alpha)= 3
\label{eq_ph38}
\end{equation}

Next, consider the 3d case, for which
\begin{equation}
\langle \cos^2 \theta \rangle = \frac{1}{3} \, , \qquad \langle \cos^4 \theta \rangle = \frac{1}{5}
\label{eq_ph39}
\end{equation}
Also in this case $\langle v \rangle =0$ and $\langle v^2 \rangle$ is given by eq. (\ref{eq_ph34}).
As regards the fourth-order moment, we have
\begin{eqnarray}
\langle v^4 \rangle  = \langle ( \alpha \, v+ \sqrt{3} \, \cos \theta)^4 \rangle =
\alpha^4 \, \langle v^4 \rangle + 6 \, \alpha^2 \, \langle v^2 \rangle + \frac{9}{5}
\label{eq_ph40}
\end{eqnarray}
and therefore,
\begin{equation}
\langle v^4 \rangle = \frac{21 \, \alpha^2 +9}{5 \, (1-\alpha^2) \, (1-\alpha^4)}
\label{eq_ph41}
\end{equation}
Consequently, the kurtosis is given by
\begin{equation}
\kappa(\alpha)= \frac{(21 \, \alpha^2 +9) \, (1-\alpha^2)}{5 \, (1-\alpha^4)} = \frac{3  \, (3+ 7 \,\alpha^2)  }{5 \, (1+\alpha^2)}
\label{eq_ph42}
\end{equation}
Also in this case, the Gaussian limit is recovered for $\alpha \rightarrow 1$.
Conversely, in the limit for $\alpha \rightarrow 0$ the kurtosis attains its minimum value $\kappa_{\rm min}$,
where
\begin{equation}
\kappa_{\rm min}=
\left \{
\begin{array}{ccc}
\frac{3}{2} \;\; & \;\;\; & d=2 \\
\frac{9}{5} \;\; & \;\;\; & d=3
\end{array}
\right .
\label{eq_43}
\end{equation}
\begin{figure}
\begin{center}
\includegraphics[width=8cm]{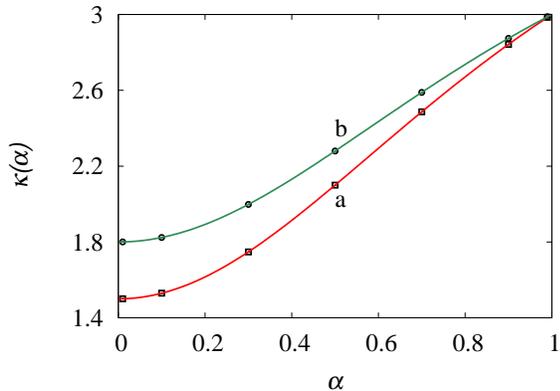}
\end{center}
\caption{Kurtosis $\kappa(\alpha)$ vs $\alpha$. Symbols are the results of stochastic simulations,
lines represent the analytical expressions eqs. (\ref{eq_ph37}), (\ref{eq_ph42}). Line (a) and ($\circ$): 2d case,
line (b) and ($\square$): 3d case.}
\label{Fig_ph1}
\end{figure}
Figure \ref{Fig_ph1} depicts the simulation results for the kurtosis compared with
the analytical predictions eqs. (\ref{eq_ph37}), (\ref{eq_ph42}). These results   refer to an ensemble of $10^9$
realizations of the process.

\begin{figure}
\begin{center}
\includegraphics[width=8cm]{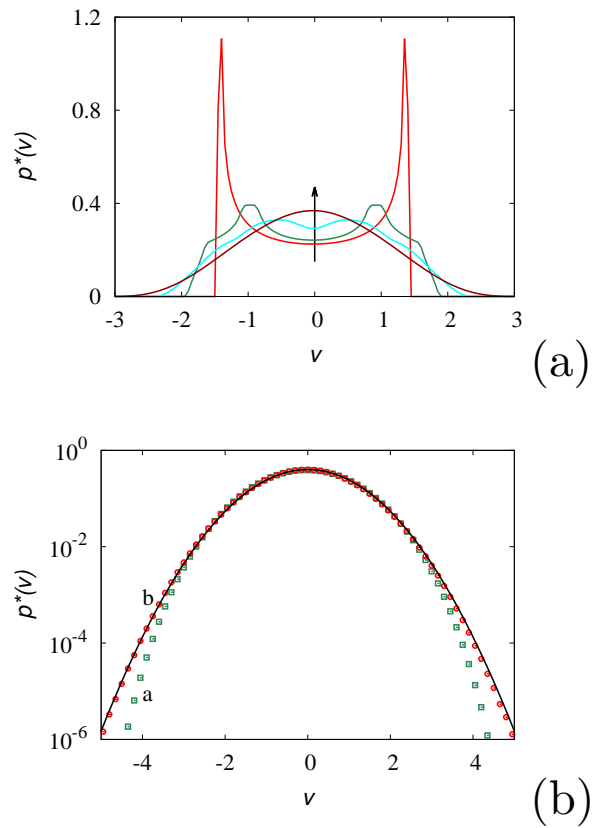}
\end{center}
\caption{Equilibrium velocity distribution $p^*(v)$ of a generic  Cartesian entry of ${\bf v}$ in the
2d case. Panel (a) refers to $\alpha=0.1,\,0.3,\, 0.5,\,0.7$.
The arrow indicates increasing values of $\alpha$.
Panel (b) refers to high values of $\alpha$: symbols ($\square$) correspond to stochastic simulations at $\alpha=0.9$,
symbols ($\circ$)   at $\alpha=0.99$.  The solid line represents the normal distribution $p_n(v)$.}
\label{Fig_ph2}
\end{figure}

\begin{figure}
\begin{center}
\includegraphics[width=8cm]{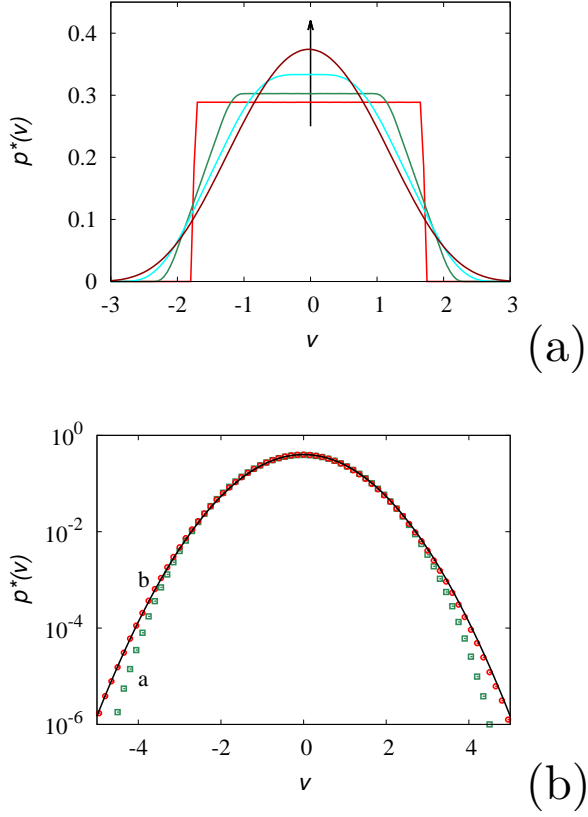}
\end{center}
\caption{Equilibrium velocity distribution $p^*(v)$ of a generic  Cartesian entry of ${\bf v}$ in the
3d case. Panel (a) refers to $\alpha=0.1,\,0.3,\,0.5,\,0.7$. The arrow indicates increasing values of $\alpha$.
Panel (b) refers to high values of $\alpha$: symbols ($\square$) correspond to stochastic simulations at $\alpha=0.9$,
symbols ($\circ$)   at $\alpha=0.99$.  The solid line
 represents the normal distribution $p_n(v)$.
}
\label{Fig_ph3}
\end{figure}

The density functions for a generic entry of the velocity field (say $v_1$ in the 2d case and $v_3$ in the
3d case) are depicted  is figures \ref{Fig_ph2} and \ref{Fig_ph3}  for the 2d and the 3d case, respectively.
The velocities appearing in these figures are normalized to unit variance.

As expected from the analysis of the kurtosis,  for low values of $\alpha$, these distributions deviate significantly
from the normal distribution $p_n(v)$,
\begin{equation}
p_n(v)= \frac{1}{\sqrt{2 \, \pi}} \, e^{-v^2/2}
\label{eq_ph44}
\end{equation}
In the limit for $\alpha \rightarrow 1$, $p^*(v)$ approaches $p_n(v)$ as 
expected. Already at $\alpha=0.99$
the resulting normalized velocity density is indistinguishable from the
normal distribution.

\begin{figure}
\begin{center}
\includegraphics[width=8cm]{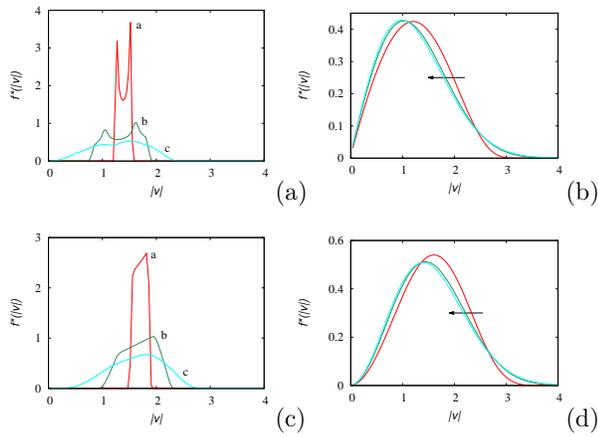}
\end{center}
\caption{Equilibrium distributions $f^*(|{\bf v}|)$ of the modulus $|{\bf v}|$ obtained from stochastic
simulations. Panels (a) and (b): 2d case, panels (c) and (d): 3d case. Lines (a) refer to $\alpha=0.1$,
lines (b) to $\alpha=0.3$, lines (c) to $\alpha=0.5$. The arrows in panels (b) and (d) indicate increasing
values of $\alpha=0.7,\, 0.9,\, 0.99$.}
\label{Fig_ph4}
\end{figure}

The equilibrium distributions   $f^*(|{\bf v}|)$ for the modulus of the normalized velocity $|{\bf v}|$ are
depicted in figure \ref{Fig_ph4} for the sake of completeness, although they do not add any 
further physical insight  to the above analysis of velocity statistics.

As regards the two asymptotic distributions obtained in the limit for $\alpha \rightarrow 1$ and $\alpha \rightarrow
0$, their mathematical justification is rather straightforward, but their
 physical interpretation  is
rather interesting.

As discussed above, the Gaussian profile is recovered in the limit for $\alpha \rightarrow 1$. In the
present case, this corresponds to the situation in which the velocity dynamics possesses the strongest memory
of its past history. The latter interpretation follows also from the exponential
decay of the velocity autocorrelation function that, 
for $\alpha \rightarrow 1$, is characterized by an 
exponent $\eta_r/m \rightarrow 0$.
In this sense the fluctuation-dissipation relation can be viewed as a dissipation-memory condition
for particle-photon interactions (momentum exchange).

In the  particle-photon dynamics described by eq.  (\ref{eq_ph1}) the above mentioned memory effects (that
should not be confused with the lack of Markovianity, as the process is strictly Markovian, and its transition mechanism has no memory) determine the
occurrence of a normal distribution for the velocity entries. This phenomenon can be easily interpreted by
considering the simplest linear relaxation dynamics for an observable $y(t)$,
\begin{equation}
\frac{d y(t)}{d t}= - \ell \,  y(t) - f(t) 
\label{eq_ph45}
\end{equation}
where $f(t)$ is a stochastic impulsive forcing and $\ell>0$ the relaxation rate,  the solution of which,
neglecting the decaying initial condition, is expressed by the convolutional integral
\begin{equation}
y(t)=  \int_0^t e^{-\ell (t-\tau) } \, f(\tau) \, d\tau
\label{eq_ph46}
\end{equation}
Assuming $f(t)=\sum_{i=1}^\infty f_i \, \delta(t-t_i^*)$, where $t_i^* < t_{i+1}^*$m    and $f_i$ generic random variables,
we have in the limit for $\ell \rightarrow 0$ that
\begin{equation}
y(t)= \sum_{i=1}^{n(t)} f_i
\label{eq_ph47}
\end{equation}
where $n(t)$ is the integer $n(t)= \sum_{i=1}^\infty \int_0^t \delta(\tau-t_i^*) \, d \tau$.

Eq. (\ref{eq_ph46}) clearly indicates that the only physical way for the velocity dynamics could perform a summation
of the  random photon momentum kicks, corresponding to the classical setting of the
Central Limit Theorem, is to possess an infinite memory of its past history, corresponding to $\ell \rightarrow 0$.
In this case, the velocity dynamics corresponds to the summation of the independent random kicks induced by the
photon bath.

\begin{figure}
\begin{center}
\includegraphics[width=8cm]{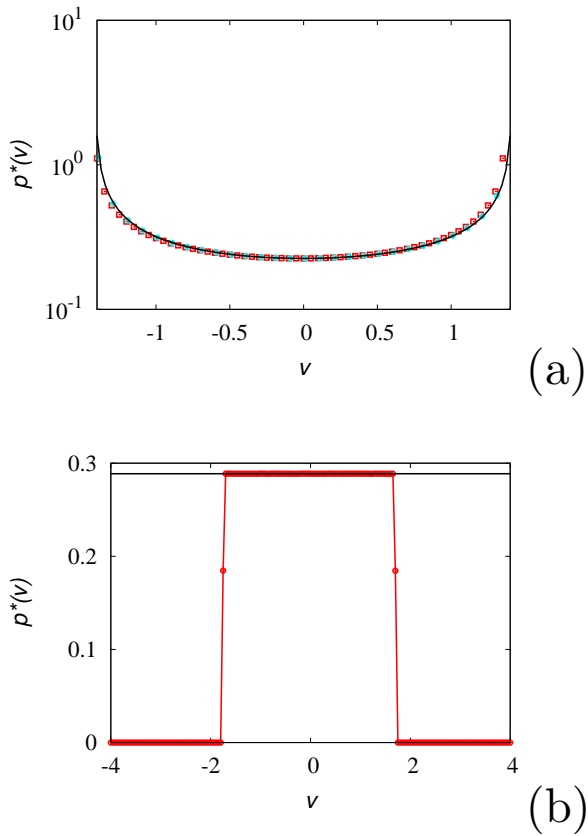}
\end{center}
\caption{Equilibrium velocity distribution $p^*(v)$ at low $\alpha$-values. Panel
(a) refers to the 2d case. Symbols represent the results of stochastic simulations at $\alpha=0.01$ ($\square$),
and $\alpha=10^{-6}$ ($\circ$). The solid line is the analytical expression eq. (\ref{eq_ph49}).
Panel (b) refers to the 3d case. Symbols  represent the results of stochastic simulations at $\alpha=0.01$.
the horizontal line is the analytical value $p^*(v)=1/2 \sqrt{3}$.}
\label{Fig_ph5}
\end{figure}

It is also interesting to consider the other limit, namely $\alpha \rightarrow 0$, corresponding to the complete absence of memory in velocity dynamics, as eq. (\ref{eq_ph11a}) reduces in this limit to
\begin{equation}
{\bf v}_{n+1}= \beta \, {\bf r}_{n+1}
\label{eq_ph47bis}
\end{equation}
and consequently, the velocity statistics is simply a  rescaled sampling of the statistics of photon momenta.
Consider the 2d case, and let $v$ a velocity entry, say  $v_1$, for which in the limit for $\alpha \rightarrow 0$
we have (upon normalization)
\begin{equation}
v= \sqrt{2} \, \cos \phi   
\label{eq_ph48}
\end{equation}
$\phi$ being uniformly distributed in $[0,2 \pi)$. The  equilibrium distribution function for $v$ is thus given by
\begin{eqnarray}
F^*(v)  & =  & \frac{1}{2 \, \pi} \int_{\{\sqrt{2} \, \cos \phi < v \}}  \, d \phi = \frac{1}{\pi} \int_{\arccos y/\sqrt{2}}^\pi d \phi \nonumber \\
& = &= 1- \frac{1}{\pi} \arccos \left ( \frac{v}{\sqrt{2}} \right )
\label{eq_ph48bis}
\end{eqnarray}
in the interval $v \in (-\sqrt{2},\sqrt{2})$, while $F^*(v)=0$ for $v < -\sqrt{2}$ and $F^*(v)=1$ for
$v > \sqrt{2}$. Differentiating $F_v^*(v)$ with respect to $v$, the density function $p^*(v)$ follows
\begin{equation}
p^*(v)= \frac{1}{\sqrt{2} \, \pi} \, \frac{1}{\sqrt{1 - v^2/2}} \, , \qquad v \in (-\sqrt{2},\sqrt{2})
\label{eq_ph49}
\end{equation}
and zero otherwise. Figure \ref{Fig_ph5}  panel (a) compares the results of stochastic simulations of the velocity
dynamics   via eq. (\ref{eq_ph11a}) at low $\alpha$-values, and the analytical expression for the limit
velocity density function  eq. (\ref{eq_ph49}) in the 2d case.
Analogously, in the 3d case, by considering $v=v_3= \sqrt{3} \, \cos \theta$, $\theta \in [0,\pi)$,
we have
\begin{equation}
F^*(v)= \frac{1}{2} \, \int_{\arccos(v/\sqrt{3})}^\pi \, \sin \theta \, d \theta =
\frac{1}{2} \, \left ( 1+ \frac{v}{\sqrt{3}} \right )
\label{eq_ph50}
\end{equation}
in $v \in (-\sqrt{3},\sqrt{3})$. Consequently, the density function $p^*(v)$ is piecewise constant in the
limit for $\alpha \rightarrow 0$,
\begin{equation}
p^*(v) =
\left \{
\begin{array}{lll}
\frac{1}{2 \, \sqrt{3}} \;\; & \;\;\; & v \in (-\sqrt{3},\sqrt{3}) \\
0 & \;  & \mbox{otherwise}
\end{array}
\right .
\label{eq_ph51}
\end{equation}
as depicted in figure \ref{Fig_ph5} panel (b).

This result is interesting from another point of view. 
The case $\alpha \rightarrow 0$, corresponds physically
to $\gamma \rightarrow \infty$, i.e. to the low-temperature limit. In these conditions, the statistics of
particle velocity corresponds to the pure sampling of the randomness in the orientation of the incoming
photons. Due to isotropy, these orientations are distributed uniformly on the surface of the unit sphere.
As can be observed from figure \ref{Fig_ph5} and from the calculations in the main text, the  shape of the equilibrium distributions
$p^*(v)$ strongly depends qualitatively on the dimension of the physical space in which photons
travel. Therefore, it is conceptually possible the experimental
determination  of the  dimension $d$, $d=2,3,\dots$, of the physical space
 from measurements of $p^*(v)$. 

\section{Concluding remarks}
\label{sec6}
This article has focused on radiative processes and their thermalization
properties in molecular systems (gases), neglecting the
influence of particle-particle collisions.
This restriction is aimed at isolating   the role of emission/absorption
processes  for understanding their peculiar features.
The interplay between radiative processes and elastic collisions
will be addressed in a forthcoming work.
Under these conditions, the equation for the
temporal evolution  of particle momentum can be expressed in the
form of a nonlinear  impulsive differential equation, driven by the
distributional derivative of a Poisson counting process.
This formulation, that accounts for the Einsteinian representation of
radiative processes, introduces the concept of a 
radiative mass $\eta$, describing
statistically the  dissipative recoil effect associated with a radiative 
transition between two energy levels. At high temperatures, the radiative
mass is smaller  than the inertial mass, while it diverges for $T \rightarrow
0$. From this formulation, a radiative
fluctuation-dissipation theorem can be derived, associated with
the exponential decay of the  velocity autocorrelation function.

The velocity distribution function has been thoroughly analyzed.
In the limit of small radiative friction, the velocity distribution  
converges to the Maxwellian (Gaussian) profile, while in the limit
of high radiative dissipation it converges to eq. (\ref{eq_ph48bis})
controlled by the random and uniform distribution of the incoming/emitted
photons. Deviations from Gaussianity are not surprinsing, as  the
momentum equation (\ref{eq_ph1}), bears some  similarity 
with the approach followed in \cite{athermal1} for the 
statistical mechanical properties
of athermal system. 
 For the sake of completeness, it should be observed that, 
while the values of $\langle v_i^2 \rangle$ are not affected by
elastic particle-particle collisions, the latter modify significantly
the shape of
the velocity distribution function.

\section*{Appendix - Nonlinear impulsive differential equations}
\label{secapp}
The analysis of differential equations of the form (\ref{eq_ph1}),(\ref{eq_ph2})
in which the impulsive forcing is modulated by a function of the
unknown variable (situation that can be referred to as a
nonlinear impulsive differential equation, in analogy with the definition
of nonlinear Langevin equations)
poses mathematical problems similar  to those encountered  for 
the nonlinear Langevin equations (driven by  a Wiener forcing),
associated with the Ito-Stratonovich dilemma.
The mathematical problems arise because the function ${\bf v}(t)$ is discontinuous at $t=t^*$ and, therefore, a rule should be specified  to intepret
this equation.
In most of the literature \cite{imp1,imp2,imp3}  a mid-point rule as been adopted. Let ${\bf v}={\bf v}(t^*_-)$ and ${\bf v}^\prime = {\bf v}(t_+^*)$,
where $t^*_\pm= \lim_{\varepsilon \rightarrow 0} {\bf v}(t^*\pm \varepsilon)$,
$\varepsilon>0$,  and integrate eq. (\ref{eq_ph2})
in the interval $[t^*-\varepsilon,t^*+\varepsilon]$ taking the limit
for $\varepsilon \rightarrow 0$,
\begin{equation}
m \, ({\bf v}^\prime- {\bf v}) = \int_{t^*_-}^{t^*_+} ( -\eta \, {\bf v}(t) + {\bf b}) \, \delta(t-t^*) \, dt
\label{eqa1}
\end{equation}
the mid-point rule assumes that  for the integral containing ${\bf v}$
\begin{equation}
\int_{t^*_-}^{t^*_+} {\bf v}(t) \, \delta(t-t^*) \, dt
= \frac{{\bf v }^\prime+ {\bf v}}{2}
\label{eqa2}
\end{equation}
so that eq. (\ref{eqa1}) becomes
\begin{equation}
m \, ({\bf v}^\prime- {\bf v})  = - \eta \, \left ( \frac{{\bf v}^\prime+{\bf v}}{2} \right ) + {\bf b}
\label{eqa3}
\end{equation}
This approach has been critically confuted in \cite{impulsive_critic}
on the basis of simple principles of calculus.
In point of fact, expressing eq. (\ref{eq_ph2}) conponentwise,
\begin{equation}
m \, \frac{d v_i}{\eta \, v_i - b_i} = - \delta(t-t^*) \, d t
\label{eqa4}
\end{equation}
$i=1,2,3$,
integrating over $(t^*_-,t_+^*)$,
\begin{equation}
\frac{m}{\eta} \int_{t_-^*}^{t_+^*} \frac{d v_i}{v_i - b_i/\eta} = -
\int_{t_-^*}^{t_+^*} \delta(t-t^*) \, d t
\label{eqa5}
\end{equation}
one finally obtains
\begin{equation}
\frac{m}{\eta} \, \log \left (\frac{v_i^\prime-b_i/\eta}{v_i-b_i/\eta}
\right ) = - 1
\label{eqa6}
\end{equation}
leading to eq. (\ref{eq_ph3}). In the limit for $\eta/m  \ll 1$, the
mid-point rule is recovered.

\end{document}